\documentclass[fp,twocolumn
]{jpsj3}
\usepackage{txfonts}
\usepackage{cite}

\title{NMR Evidences for the coupling between conduction electrons and molecular degrees of freedom in the exotic member of the Bechgaard salt, (TMTSF)$_2$FSO$_3$	
}

\author{Hidetaka SATSUKAWA$^1$, Akio YAJIMA$^1$, Ko-ichi HIRAKI$^1$\thanks{ko-ichi.hiraki@gakushuin.ac.jp}, Toshihiro TAKAHASHI$^1$, 
Haeyong KANG$^2$, 
 Younjung JO$^3$, 
  Woun KANG$^4$ and Ok-Hee CHUNG$^5$}
\inst{$^1$Department of Physics, Gakushuin University, Tokyo 171-8588, Japan \\
$^2$Department of Energy Science, Sungkyunkwan University, Suwon, 16419, Korea\\
$^3$Department of Physics, Kyungpook National University, Daegu, 41566, Korea\\
$^4$Department of Physics, Ewha Womans University, Seoul 03760, Korea \\ %
$^5$Department of Physics Education, Sunchon University, Suncheon 57922, Korea %
} 

\abst{
We performed $^{77}$Se and $^{19}$F-NMR measurements on single crystals of 
(TMTSF)$_2$FSO$_3$ to characterize the electronic structures of 
 different phases 
 in the Temperature-Pressure phase diagram, 
determined by precise transport measurements 
[Jo \textit{et al}., Phys. Rev. \textbf{B67}, 014516 (2003)].
We claim that such varieties of electronic states %
 in the refined phase diagram 
are 	
caused by strong couplings of the conduction electrons with 
FSO$_{3}$ anions, 
especially  with the permanent electric dipoles on the anions.  
We suggest that 
as temperature decreases, the FSO$_3$ anions form orientational ordering through two steps; 
first only the tetrahedrons form an orientational order leaving the orientations of the electronic dipoles in random (transition I); 	
then the dipoles form a perfect orientational order at a lower temperature (transition II). 
In the intermediate temperature range between transitions I and II, 
we found an appreciable enhancement of homogeneous and inhomogeneous widths of  $^{77}$Se-NMR spectrum.  
From the analysis of the angular dependence of the linewidth, 
we attributed these anomalies to the intramolecular charge 
disproportionation or imbalance 		
and its slow dynamics caused by the coupling with the permanent electric dipole of anion. 
Results of $^{19}$F-NMR relaxation and lineshape measurements support this picture very well.  
Electronic structures at higher pressures up to  1.25~GPa are 
discussed on the basis of the results of the $^{77}$Se and $^{19}$F-NMR measurements.  
}

\kword{(TMTSF)$_2$FSO$_3$, anion dynamics, NMR
}

\begin{document}
\maketitle

\section{Introduction}
\begin{figure}[htbp]
\begin{center}
\includegraphics[width=6cm]{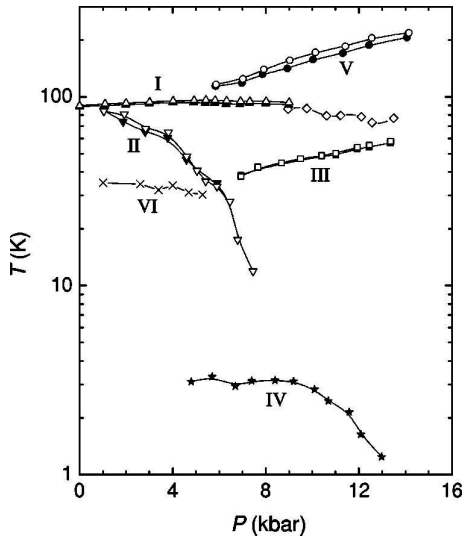} 
\caption{
Temperature-Pressure phase diagram of (TMTSF)$_2$FSO$_3$, indicating transitions I to VI.	
Reproduction of 		
Fig. 6 of 
ref. \ref{bib:jo2003}.		
Pressure values are indicated in unit of kbar: 10~kbar = 1~GPa.%
}
\label{fig:phaseDiagram_0}
\end{center}
\end{figure}

The title compound is an exotic member of the well-known Bechgaard salts:  
a charge transfer complex made of an organic TMTSF and a tetrahedral counter anion, FSO$_{3}^{-}$, 
carrying a permanent electric dipole moment.
This 	salt 	was synthesized to study the effect of the introduction of asymmetric 
dipolar anions on the conducting electron systems.\cite{wudl1982} 	
Several studies were done to clarify the electronic properties of this salt in 1980's,\cite{lacoe1983,gross1984}
however, no clear experimental evidence for the effects of permanent dipoles was observed.	
Twenty years later, precise transport measurements were performed for high quality single crystal samples 
and the temperature-pressure ($T$-$P$) phase diagram was refined.\cite{jo2003,kang2003}
In addition to the basic structures earlier reported,  
several novel anomalies in resistivity and 
 thermoelectric power 
 were detected under pressure.	
The authors proposed new phase boundaries from the transitions I to VI as shown in Fig.~\ref{fig:phaseDiagram_0}. 
The complexity of the proposed phase diagram seemed to indicate the contributions	
of some novel degrees of freedom, probably due to the introduction of electric dipoles in this salt. 
To address this issue from microscopic points of view, 
we performed %
 NMR measurements. 

We have already reported some of the results of $^{77}$Se-NMR study at ambient pressure 
as well as under pressure of 0.65~GPa using a conventional 
clamp type pressure cell.\cite{hiraki2003} 
We found several interesting features; i) the insulating state below the 
metal-insulator transition at ambient pressure is nonmagnetic with a finite spin gap;\cite{hiraki2003} 
ii) an appreciable line broadening was observed in the region between transitions I and II under pressure; 
iii) below transition II, the system turns into a nonmagnetic insulating state; 	
iv) the spin gap in the insulating state decreases with pressure.

In the present study, we performed several additional $^{77}$Se and $^{19}$F-NMR measurements 
using 
 single crystal samples to clarify the more detailed electronic structure and the roles of anion dynamics under pressure. 	
$^{77}$Se measurements were performed in a different geometry from the previous works\cite{satsukawa2005,takahashi2005}.	 

In this report, we first describe the earlier NMR data at ambient pressure and 0.65~GPa 
in \S \ref{sec:ap} and \S \ref{sec:press8p1}, respectively, 
which we could not show in detail in our previous reports\cite{hiraki2003,satsukawa2005,takahashi2005}. 
New results of $^{77}$Se-NMR measurements at 0.65~GPa and 1.25~GPa are shown 
in \S \ref{sec:press8sat} and  \S \ref{sec:press11}, respectively. 
Results of $^{19}$F-NMR measurements at ambient pressure, 0.45~GPa and 0.9~GPa are described in \S \ref{sec:F-ap}, 
\S \ref{sec:F-0.45} and  \S \ref{sec:F-0.9}, respectively. 
Finally, possible electronic structures and the role of anion dynamics are discussed in \S \ref{sec:conc}.

\section{Experimental details}
$^{77}$Se and $^{19}$F-NMR spectra and relaxation rates were measured using single crystal samples 
with a typical size of $	
1.9\times 0.9\times0.5$ mm$^3$.
$^{77}$Se experiments were 
performed at an applied magnetic field 
of 		
$H_0$	
$\sim$ 		
9~T,  
which corresponds to the resonance frequency of (2$\pi$)$\times73$ MHz (=$^{77}\gamma$$H_0$, 
where $^{77}\gamma/(2\pi)=8.131$ MHz/T is the gyromagnetic ratio of $^{77}$Se nuclei). 
The magnetic field strength was calibrated by the $^{13}$C-NMR signal of a standard sample of tetramethylsilan 
(TMS = Si(CH$_3$)$_4$).\cite{nmrref}
NMR spectra were obtained by the fast Fourier transformation (FFT) of the spin echo signals following $\pi$/2-$\pi$ 
radio frequency (\textit{rf})  pulse sequences. 
Linewidh of the spectrum was defined as the square root of the second moment.

Spin-lattice relaxation rate, $T_1^{-1}$, was measured from the recovery of the nuclear magnetization 
obtained by $\pi$ (or saturation comb pulses)-$\tau$-$\pi$/2-$\tau$'-$\pi$ pulse sequences.  
The echo decay rate, $T_2^{-1}$, or the ``homogeneous" linewidth was also determined by 
measuring the decay of the echo intensities as a function of $\tau$' with a fixed $\tau$.
The linewidh, defined as the square root of the second moment of the spectrum, 
corresponds to the total transverse relaxation rate, $T_2^{*-1}$, including ``inhomogeneous" linewidth.

Hydrostatic pressure was applied with the use of clamp-type 
pressure cells made of Be-Cu alloy, 
in which an NMR coil containing a single crystal sample was set with Daphne 7373 as a pressure medium.
We used two types of pressure cell.
$^{77}$Se-NMR measurements described in \S\ref{sec:press8p1} 
and all $^{19}$F-NMR measurements were done with a conventional type cell with sizes of 25 mm in diameter and 87 mm in length. 
This cell was too long to rotate in the cryostat. 	
Therefore we did not change the field direction with this cell. 
External field	
was 		
applied almost perpendicular to the $a$ axis. 
Another type of pressure cell with sizes of 25mm in diameter and 40 mm in length was 
a hand made one designed by one of the authors (WK).  
With the use of this ``mini" cell it became possible to mount the sample to a goniometer and rotate it in the external field.\cite{dimension}

When we applied a pressure of 0.8~GPa at room temperature, 	
the actual pressure at low temperatures below 150~K was estimated as $\sim$ 0.65~GPa,
from an independent resistance measurement on a small piece of crystal with the same pressure medium; 
the pressure drop at low temperatures due to the thermal contraction was estimated as about 0.15~GPa.  
This agrees well 
with 	
the reported properties of Daphne 7373;~\cite{Daphne7373} 
it was claimed that the pressure reduction at low temperature from room temperature is  always 0.15~GPa, 
irrespective the initial pressure at room temperature, at least up to 1.5~GPa.~\cite{Daphne7373}  
The pressure values mentioned in the text are those expected at low temperatures below 150~K, 
applying this correction, unless otherwise noted.  
		
$^{77}$Se-NMR measurements 
with the ``mini" cell 
were performed 		
in a geometrical configuration, 
where the NMR coil was wound along the $b'$ axis and the external field was rotated in the $ac^*$ plane.
At the Se-sites, 4p$_\mathrm{z}$ orbital is considered to give a dominant 	contribution 
to the conducting HOMO band, so that the hyperfine coupling between the conduction electrons 
and the Se-nuclei is expected to have an uniaxial symmetry. 	
Since the principal axis of the p$_\mathrm{z}$ orbital is almost parallel to the $a$ axis,\cite{zhang2005}
in the present experimental geometry, one can detect the parallel and perpendicular components of the Knight shift tensor, 
when the field is parallel to the $a$ and $c^*$ axes, respectively. 
The Knight shift, $K$, is expressed as 
$K=K_\mathrm{iso}+K_\mathrm{aniso}$, where $K_\mathrm{iso}$ and $K_\mathrm{aniso}$ are 
the isotropic and anisotropic components 	of the Knight shift, respectively.	
For an uniaxial case, $K_\mathrm{aniso}$ is written as  %
$K_\mathrm{aniso}=K_\mathrm{ax}(3\cos^{2}\theta-1)$,	%
where $\theta$ is the angle of the applied field measured from the p$_\mathrm{z}$ orbital. %
Since $K$ is given as 
$K(\theta)=A(\theta)\chi_\mathrm{s}=(A_\mathrm{iso}+A_\mathrm{aniso}(\theta))\chi_\mathrm{s}$, 
using the hyperfine coupling constant, $A(\theta)$, and the local spin susceptibility of electrons	per molecule, $\chi_\mathrm{s}$,	
the measurements of the Knight shift enable us to detect the local susceptibility	at individual molecular sites. 
In the case of 2:1 charge transfer complexes, (with homogeneous spin/charge distributions), 
$\chi_\mathrm{s}$ is given as $\chi_\mathrm{s}=\chi_\mathrm{mol}/2N_\mathrm{A}\mu_\mathrm{B}$ 
using the molar susceptibility, $\chi_\mathrm{mol}$, 	the Avogadro's number, $N_\mathrm{A}$, 	and the Bohr magneton, $\mu_\mathrm{B}$.
The factor 2 comes from the fact that the unit cell contains two TMTSF molecules. 
(The reason to normalize $\chi_\mathrm{s}$ with $\mu_\mathrm{B}$ is because the hyperfine coupling constant, $A$, 
is  usually given in the unit of T/$\mu_\mathrm{B}$) 	

The $^{19}$F-NMR measurements were carried out 	at applied magnetic fields of 3 $\sim$ 6~T
(the gyromagnetic ratio of $^{19}$F nuclei	is $^{19}\gamma/(2\pi)=$ 40.05 MHz/T).
We had carefully chosen fluorine free materials in applying pressure. 
(We gave up to use the commonly used pressure capsule made of teflon but used  the one made of polyvynil chloride.)	
The geometrical situations  in the $^{19}$F measurements were completely 
the same as in our earlier reports\cite{satsukawa2005,takahashi2005}, 
that is, the magnetic field was applied perpendicular to the one-dimensional conducting $a$ axis.

\section{Results and Discussions}
%
%
%
\subsection{\label{sec:ap}$^{77}$Se-NMR at ambient pressure}
\begin{figure}[htbp]
\begin{center}
\includegraphics[width=8cm]{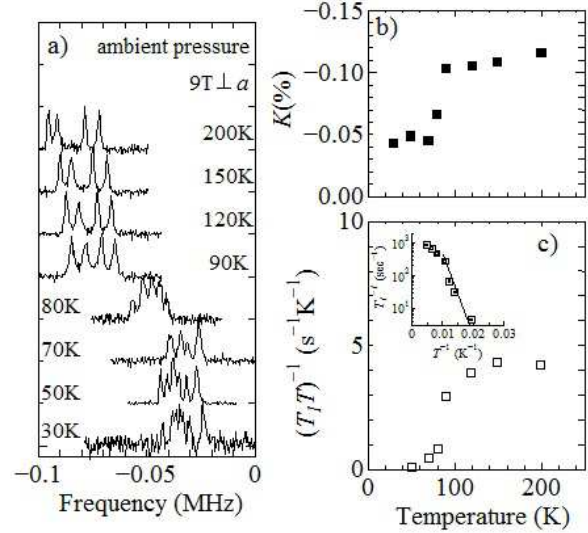}	
\caption{Summary of the $^{77}$Se NMR results at ambient pressure; 	
a) NMR spectra at several temperatures and 
b) the temperature dependences of  the central  position of the spectrum and c) $(T_1T)^{-1}$.	
The origin of the horizontal axis in a) is $^{77}\gamma H_0/(2\pi)$.  	
The inset of c) shows a semi-logarithmic plot of $T_1^{-1}$ as a function of inverse temperature.
The solid line of the inset indicates 
$T_1^{-1} \propto \exp(-\Delta/k_\mathrm{B}T)$ with $\Delta/k_\mathrm{B} \sim$ 420 K.
}
\label{fig:vspap}
\end{center}
\end{figure}

The results of $^{77}$Se-NMR lineshape, spectral shift and spin-lattice relaxation rate, $T_{1}^{-1}$, 
at ambient pressure were summarized in Fig. \ref{fig:vspap}. 	
External 		%
field of 9~T was applied perpendicular to the conducting $a$-axis.	 
In the metallic state, the observed spectrum consists of four peaks,	
corresponding to the number of inequivalent Se sites 
in a unit cell (eight Se sites are reduced to four because of the inversion symmetry of the crystal structure).  
The spectral position is almost temperature independent and 
$T_{1}^{-1}$ seems to follow the usual Korringa behavior, $(T_{1}T)^{-1} \sim$ %
4~s$^{-1}$K$^{-1}$. 

Drastic changes were observed at about 90~K in the spectral shape, Knight shift and $T_1^{-1}$, as shown in Fig. \ref{fig:vspap}: 
An abrupt shift of the spectral position are observed at this temperature 
and the number of the peaks increases to six or more at lower temperatures. 
$T_1^{-1}$ starts to decrease exponentially below 90 K.		
All these NMR behaviors are consistent with those expected for the metal-insulator transition due to the anion ordering (AO) 
observed at this temperature. 
$X$ ray experiments\cite{yamaura} reported the occurrence of AO with a wave number of (1/2 1/2 1/2) at this temperature. 
The increase in  the number of NMR peaks corresponds to the loss of the inversion symmetry of the crystal. 
The abrupt shift of the spectral position is due to the vanishing of %
Knight shift and the exponential reduction of $T_1^{-1}$ 
clearly indicates the opening of a spin gap at the Fermi surface (FS). 		%
The spin gap, $\Delta/k_\mathrm{B}$, was estimated as $\sim$ 420~K from the temperature dependence of $T_1^{-1}$ 	
shown in the inset of Fig. \ref{fig:vspap}~c).
This value is quite large as an anion ordering gap of TMTSF family\cite{AO_gap}.	
This may come from the fact that the tetrahedral FSO$_3$ anion carries a permanent electrical dipolar moment.
The finite frequency shift %
($\sim$ -0.04\%	) 	
remaining	 at low temperatures 	should be attributed to 
the chemical shift of the TMTSF molecule.

%
\subsection{\label{sec:press8}$^{77}$Se-NMR under pressure: 0.65~GPa} 
\subsubsection{\label{sec:press8p1}Temperature dependences for $H_{0} \perp$ $a$-axis} 
\begin{figure}[htbp]
\begin{center}
\includegraphics[width=8cm
]{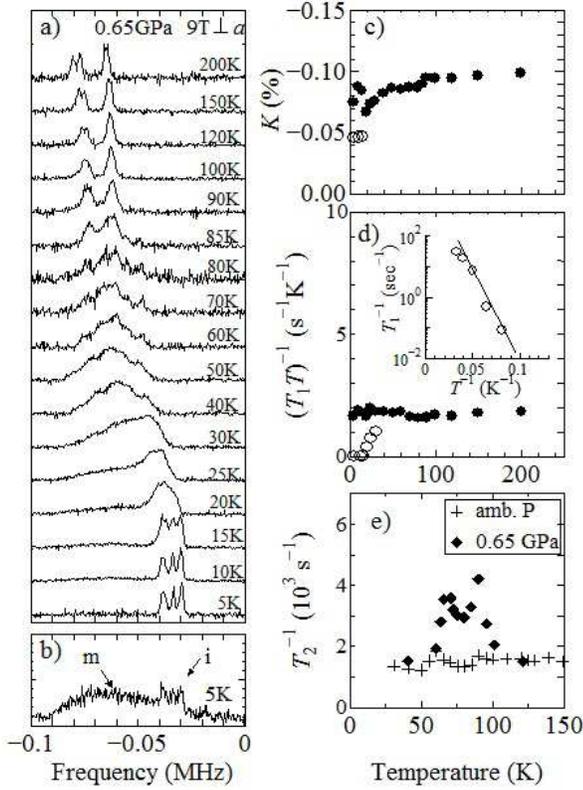}		%
\caption{Summary of the $^{77}$Se NMR results under pressure of 0.65~GPa using a conventional pressure cell;	%
a) NMR spectra at several temperatures, b) a spectrum observed  at 5 K  with a shorter repetition period	%
of $\sim$ 3~s, 	%
c) the temperature dependences of the central position of the spectrum, 
d) $(T_1T)^{-1}$ and e) the homogeneous width determined as the echo decay rate, $T_2^{-1}$,	%
at 0.65~GPa (closed diamonds) and at ambient pressure (crosses). 
The origin of the horizontal axis in a) is $^{77}\gamma H_0/(2\pi)$.  
The open circles in 
c) and 			%
d) indicate the results for the narrower component of the spectrum.  
The inset of d) 	shows the $T_1^{-1}$ as a function of inverse temperature.
The solid line of the inset indicates the $T_1^{-1} \propto \exp(-\Delta/k_\mathrm{B}T)$ 
with $\Delta/k_\mathrm{B} \sim$ 130 K.
}
\label{fig:sum8kbar}
\end{center}
\end{figure} 

Figure \ref{fig:sum8kbar} shows the results of the $^{77}$Se-NMR measurements under pressure of 0.65~GPa 
with the same experimental geometry as at ambient pressure.
The 		
spectra and relaxation behaviours %
in the metallic state at high temperatures 
were basically similar to those at ambient pressure. 		%
Knight shift, %
$K\sim$4.7$\times$10$^{-4}$, 		%
and Korringa constant, %
$(T_1T)^{-1}\sim$1.8~s$^{-1}$K$^{-1}$, 			%
were slightly reduced by applying pressure, 
as seen by comparing the results in Figs.~\ref{fig:vspap}~b) c) and Figs.~\ref{fig:sum8kbar}~c) d). %
This is natural because applying pressure generally tends to broaden the band width, 
leading to the reduction of the density of states at the Fermi level. 
Actually, the value of $(T_1TK^{2})^{-1} \sim$ 8$\times$10$^6$~s$^{-1}$K$^{-1}$ is almost the same as the value at ambient pressure.	

As temperature decreases, remarkable broadening of spectrum appears below $\sim$90~K, 
and at the same time the center of gravity of the spectrum starts to shift gradually towards the higher frequency side. 
Below $\sim$30~K, a narrower component of spectrum appears, 
and below $\sim$15~K, several sharp peaks around the zero Knight shift position coexist 
with a broad component in the lower frequency, as shown in Fig. \ref{fig:sum8kbar}~b).

Spin-lattice relaxation measurements revealed that there exist two components of relaxation at low temperatures below $\sim$ 30 K, 
as shown by open and closed circles in Fig.~\ref{fig:sum8kbar}~d), reflecting the coexistence of two components of the spectrum.   	
The relaxation rate for the broader component seen around $\sim$$-$0.07~MHz follows the Korringa-like behaviour, 
$(T_1T)^{-1}$ $\sim$ constant, as shown by closed circles, in the whole temperature range.  
This clearly indicates that the broad component comes from a metallic region of the sample in spite of the unusual broadening.
On the contrary, the relaxation rate for the sharp peaks with the smaller shift decreased exponentially with decreasing temperature, 		
as shown by open circles	
in Fig. \ref{fig:sum8kbar}~d).			%
			The energy gap $\Delta/k_\mathrm{B}$ was estimated as $\sim$130~K, 
which is much smaller than $\sim$420~K	observed in the insulating state at ambient pressure.

The temperature of $\sim$90~K, at which the line broadening and the peak shift starts to appear, 
corresponds to transition I given in the $T$-$P$ phase diagram (Fig.\ref{fig:phaseDiagram_0}).	 
It is also reasonable to consider the temperature $\sim$30~K, below which the sharp component of spectrum appears, 
as transition II, since the NMR properties at lower temperatures are quite similar to those 
in the nonmagnetic insulating state at ambient pressure. 
We note that the volume fraction of the nonmagnetic phase is much larger than that of the metallic phase.
The small fraction of the metallic phase suggests that the admixture of the metallic component 	comes 
either from some inhomogeneity in applied pressure, or from the first-order nature of transition II.

To get 
 further insight into the origin of the unusual broadening of the $^{77}$Se-NMR spectrum, 
we measured the temperature dependence of the homogeneous linewidth, $T_2^{-1}$, %
at ambient pressure and under pressure; 
the results are shown by crosses and closed %
diamonds in Fig.~\ref{fig:sum8kbar}(e), respectively.
While $T_2^{-1}$ at ambient pressure was almost constant in temperature, $\sim$1500~s$^{-1}$, 
two peak anomalies in $T_2^{-1}$ were observed at 0.65~GPa around 90~K and 70~K. 
The first peak at 90~K should be related to transition I, where most of the NMR properties have shown anomalies. 
The second peak around 70~K appears, however, in the intermediate temperature range between transitions I and II. 

We note that the appearance of the inhomogeneous broadening is accompanied 
by the enhancement of the homogeneous width and the latter shows a peak behavior. 
Similar enhancement of $T_2^{-1}$ was observed in a two dimensional system with	charge disproportionation (CD), 
$\theta$-ET$_{2}${\it M}Zn(SCN)$_{4}$, {\it M}=Rb and Cs,\cite{chiba2004} 
where extremely slow fluctuations of hyperfine fields due to the dynamics of CD were shown 
to enhance the transverse relaxation rate, $T_2^{-1}$. 
We expect that a similar situation appears in the present system below 90~K: 
Assume that the local fields responsible for the inhomogeneous broadening rapidly vary in time 
with a characteristic time, $\tau_\mathrm{c}$: 
At high temperatures, where $\tau_\mathrm{c}$ is sufficiently short, the homogeneous width should be given as 
$T_2^{-1} \sim \langle\Delta\omega^{2}\rangle\tau_\mathrm{c}$, 
where $\langle\Delta\omega^{2}\rangle$ is the second moment  due to the inhomogeneous broadening. 
At low temperatures, on the contrary, the homogeneous width should be determined by the life time of the Zeeman levels, 
as $T_2^{-1} \sim \tau_\mathrm{c}^{-1}$; 
the maximum of $T_2^{-1}$ thus appears when the condition, $\langle\Delta\omega^{2}\rangle^{-1/2}\tau_\mathrm{c}\sim 1$, is satisfied.
We will discuss the possible origin of the inhomogeneous local fields and their dynamics in the present system later.

A superconducting (SC) state was reported in this pressure region;\cite{gross1984}  
the transition temperature was $\sim$3 K which is much higher than those of the other TMTSF salts.
However the SC phase was not considered to be a bulk one; 
the reported volume fraction of SC was $\sim$ 2 \%. 
In fact, we did not observe any anomaly relating to SC in the NMR measurements down to 1.5 K. 
Therefore we do not touch the superconducting phase in the present study.

\subsubsection{\label{sec:press8sat}Angular dependence of $^{77}$Se-NMR spectrum for $H_{0}$ in ac*-plane}
\begin{figure}[htbp]
\begin{center}
\includegraphics[width=8cm
]{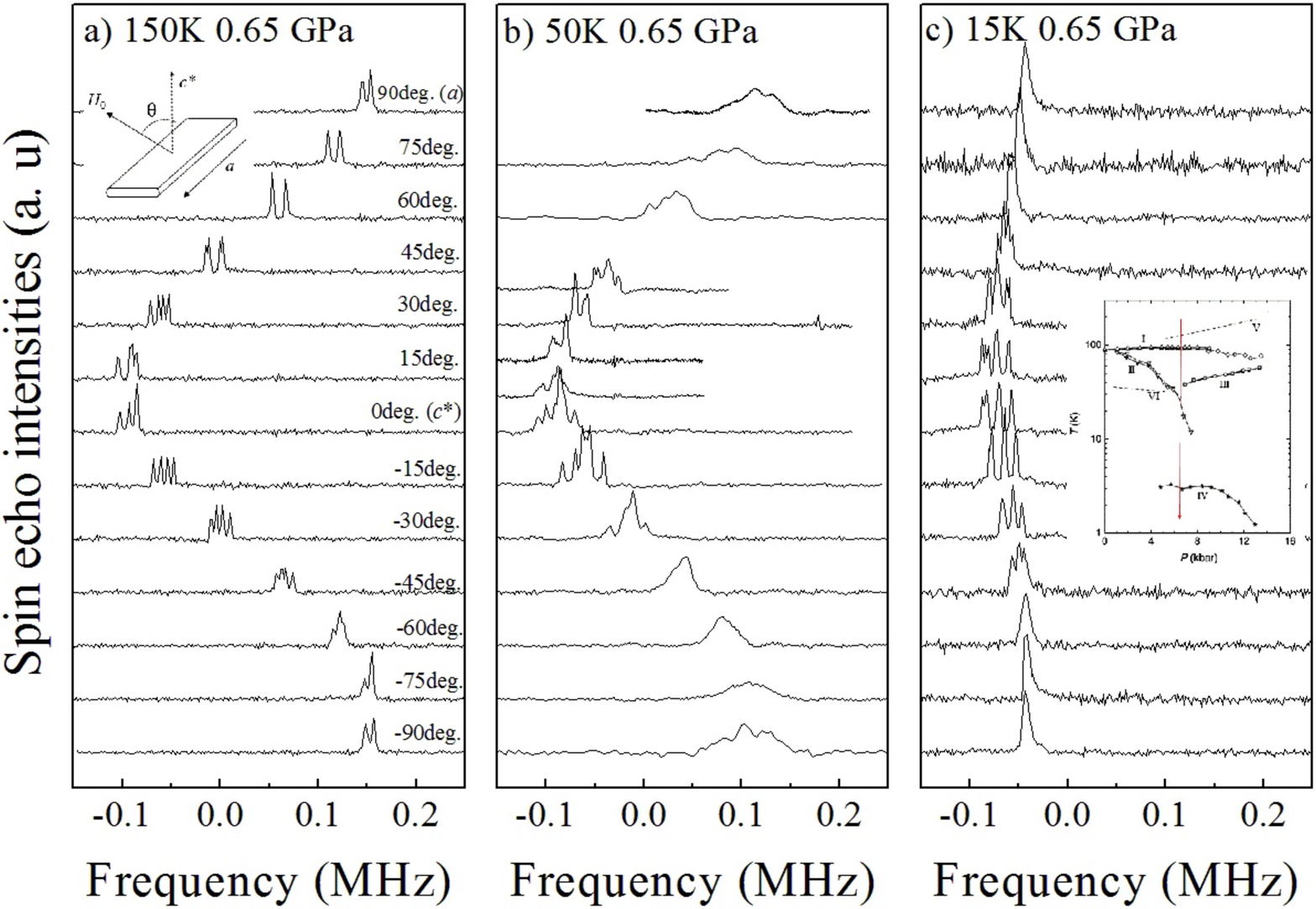} 
\caption{Angular dependence of $^{77}$Se NMR spectrum at 0.65~GPa,  
measured at temperatures of a) 150 K, b) 50 K and c) 15 K, respectively.
The origins of the vertical axes are $^{77}\gamma H_0/(2\pi)$.  
The experimental geometry is shown in the inset of a).
The inset of c) is the $T$-$P$ phase diagram, where the solid line indicates the approximate cooling path in the present measurements.
}	
\label{fig:vsp8kbar}
\end{center}
\end{figure}

In order to clarify the origin of the unusual line broadening in the region between transitions I and II, 
we investigated the angular dependence of  $^{77}$Se-NMR for the external field, $H_{0}$, rotated in the $ac^*$ plane. 
The measurements were performed at a pressure of 0.65~GPa using the	``mini'' pressure cell, mentioned before, 
to rotate in the limited sample space.
Results at 150 K (metallic state), 50 K and 15 K (insulating state) are shown in Fig. \ref{fig:vsp8kbar}~a), b) and c), respectively. 
In this configuration, the field was rotated in the plane including	the directions parallel and perpendicular to the p$_z$ orbital,	
where the anisotropy of the Knight shift is expected to be the largest. 

As shown in Fig. \ref{fig:vsp8kbar}~a), the spectrum at 150~K consists of at most four absorption peaks, 
and the center of gravity of the spectrum shows a large angular dependence characteristic of the Knight shift 
for the p$_z$ orbital at the Se site. 
The Knight shift due to the	 hyperfine coupling is positive and the largest when the external field is parallel to the p$_z$ orbital 
($H_0\parallel$ $a$ axis).	
The amplitude of the angular oscillation  is $\sim$0.25 MHz, corresponding to $K_\mathrm{ax} \sim$ 0.099\%. 	

The spectrum at 50~K is shown in Fig. \ref{fig:vsp8kbar}~b). 
One can easily find that the spectrum is broadened remarkably, while 
the angular %
variation 
of the shift was almost as large as that at 150 K.
The latter means that the average spin susceptibility does not change so much, 
indicating that the system remains metallic at this temperature in spite of the unusual broadening. 
We also note that the excess broadening is strongly angular dependent; 
the spectrum becomes broader when the shift becomes larger. 

Fig. \ref{fig:vsp8kbar}~c) shows the NMR spectra at 15~K.
The angular dependence of the spectral center	is much smaller than that at higher temperatures.
This is quite natural because the system is nonmagnetic at this temperature 
and thus the spin susceptibility is vanished. 
The angular dependence of the spectral center	is thus considered to be due to 	the anisotropy in the chemical shift of TMTSF molecule, 
and gives an unambiguous zero Knight shift position at each field orientation. 	

\begin{figure}[htbp]
\begin{center}
\includegraphics[width=8cm
]{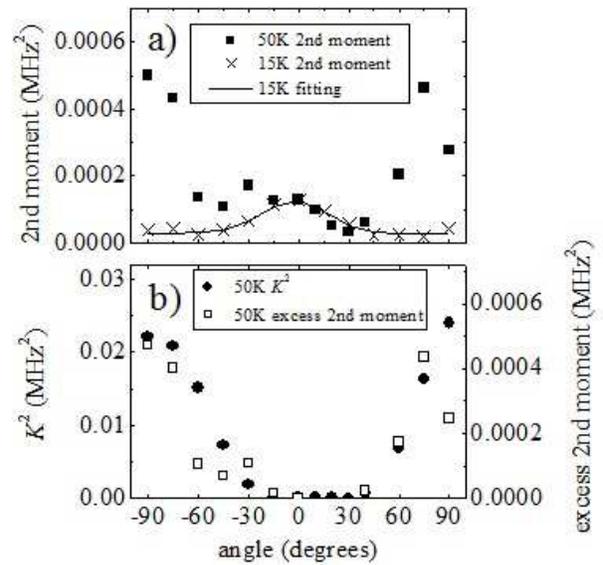} 
\caption{
a) Angular dependence of the second moment of $^{77}$Se NMR spectrum at 0.65~GPa,	 
measured at 50 K (closed squares) and 15 K (crosses). 
b) Comparison between the square of the Knight shift, $K^2$ (closed circles), and the excess second moment (open squares).
The Knight shift, $K$, was measured from the spectral center (the first moment of the spectrum) observed at 15 K for each filed direction.
}
\label{fig:8kbaranalysis}
\end{center}
\end{figure}

In order to analyze the angular dependence of the excess broadening at 50~K, 
we calculated the second moment of the spectra observed at 50~K and 15~K. 
The results are shown in Fig. \ref{fig:8kbaranalysis}~a).
The broadening or the structure %
 of the spectrum observed at 15~K is caused mainly by the difference in chemical shift
at different nuclear sites, which should be naturally included in the second moment at 50~K. 
The excess second moment at 50~K is obtained simply by subtracting the results at 15~K from those at 50~K
 (open squares in in Fig. \ref{fig:8kbaranalysis}~b)).

We found that the excess second moment scales well to the square of the spectral shift, $K^2$, 
(closed circles in Fig. \ref{fig:8kbaranalysis}~b)).
This strongly suggests that the excess broadening is caused by the inhomogeneity in the Knight shift, that is, 
by that in 	the local susceptibility. 	
Considering that the system does not contain any localized magnetic spins and remains metallic in this temperature range, 
one of possible explanations of the inhomogeneity in the local susceptibility may be intermolecular %
CD in the chains, 
as confirmed in some two dimensional systems, such as $\theta$-(ET)$_2$RbZn(SCN)$_4$\cite{chiba2004} 
and $\lambda$-(BETS)$_2M$Cl$_4$.\cite{hiraki2007,hiraki2010}
However, this does not seem to be the case because of the following reasons; 
i) there is no experimental evidence for CD in the TMTSF salts, while some TMTTF salts do exhibit charge ordering 
(CO)\cite{brown2002,fujiyama2006,hirose2010} %
leading to nonmetallic conductivity. 
ii) the inhomogeneous broadening is rather small or moderate compared with the one observed in the two dimensional systems\cite{chiba2004, hiraki2007, hiraki2010}.

We estimate the degree of the disproportionation of the local susceptibility, $\Delta\chi/\chi$,  
in this salt as we did on the other systems.\cite{chiba2004,hiraki2010} 
Since the angular dependence of $K^2$ and the excess second moment ($\Delta K^2$) are well scaled with each other 
as demonstrated in Fig. \ref{fig:8kbaranalysis}b), we obtain an estimate of $\Delta K/K\sim\Delta\chi/\chi\sim$ 0.14. 
If one assumes that the local susceptibility is proportional to the local charge and the average valence of TMTSF is 0.5, 
the distribution of the molecular charge is estimated as $\sim0.5\pm 0.07$. 
The degree of CD is much smaller than in the TMTTF family \cite{brown2002
,fujiyama2006} 
and the other CD systems.\cite{chiba2004,hiraki2010
} 

Another possibility of the inhomogeneous broadening is an intramolecular charge/spin 
disproportionation or 		%
imbalance 
caused by possible couplings with the permanent electric moments of anion. 	
In this case, it seems natural that the dynamics of the local fields at the Se sites, 
given by the hyperfine couplings with the conduction electron spins,  
should be strongly coupled with the dynamics of anions or the reorientation of the permanent dipoles. 
At high temperatures, where the anions are rotating in a three-dimensional (3D) manner, 
the fluctuation of the local field should be too rapid to contribute to $T_2^{-1}$. 
Below transition I, one can assume that the CD (of the order of $\sim0.5\pm0.07$) may appear because of partial AO, 
but it may be still rapidly changing in time with a characteristic frequency, $\tau_\mathrm{c}^{-1}(T)$,   
because of hindered motions of anions.    
The line width due to the inhomogeneity of the local fields should be motional-narrowed as given before.
This picture seems to explain the observed results of the $^{77}$Se-NMR measurements very well. 
The confirmation of this picture will be given by $^{19}$F-NMR in the later sections.

%
\subsection{\label{sec:press11}$^{77}$Se-NMR under pressure: 1.25~GPa} 

\begin{figure}[htbp]
\begin{center}
\includegraphics[width=8cm
]{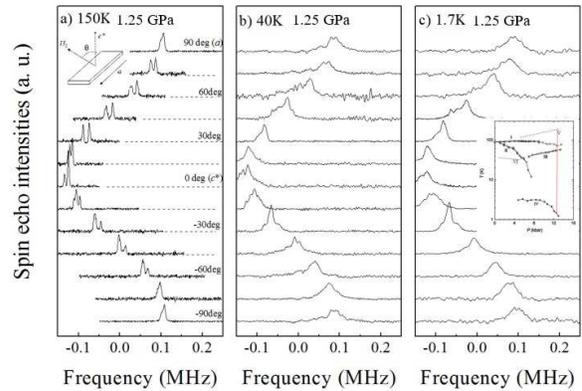}	%
\caption{$^{77}$Se NMR spectra under 1.25~GPa at a) 150 K, b) 40~K and c) 1.7~K, respectively.
The solid line in the inset of c) shows the approximate cooling path in the phase diagram. 
}
\label{fig:vsp14kbar}
\end{center}
\end{figure}

To investigate the metallic state in detail, we applied 
high pressure up to 1.25 GPa at low temperatures below 150 K %
Angular dependence of $^{77}$Se-NMR spectrum at this pressure was measured at several temperatures, 
as shown in Fig. \ref{fig:vsp14kbar}.
We note that the results at 150~K is quite similar to those observed at 0.65~GPa, apart from a slight  excess broadening: 
The amplitude of the angular dependence of the spectral shift, $K_\mathrm{ax}$, is estimated as 0.091 \% at 1.25~GPa, 
which is comparable to the value of 0.099\% at  0.65~GPa. 
(Here, the contributions of chemical shift were corrected, using the data at 15~K at 0.65~GPa.)	
Temperature dependence of $K_\mathrm{ax}$ is rather small down to 1.7~K at this pressure; 
$K_\mathrm{ax}$ is 0.085 \%  and 0.08\% at 40~K and 1.7~K, respectively. 
Since the 	Knight shift is proportional to the spin susceptibility, $\chi_\mathrm{s}$, as mentioned before, 
this result is consistent to 	the Pauli-like behavior of  $\chi_\mathrm{s}$, expected in a simple metal without electron correlations.

The inhomogeneous line broadening at low temperatures was also observed at this pressure. 
Considering the broadening to be due to the inhomogeneity in the Knight shift, we obtained $\Delta K/K \sim$ 0.3 at 40~K 
and 0.39 at 1.7~K at this pressure; the inhomogeneity increases as temperature decreases. 	

\begin{figure}[htbp]
\begin{center}
\includegraphics[width=6cm
]{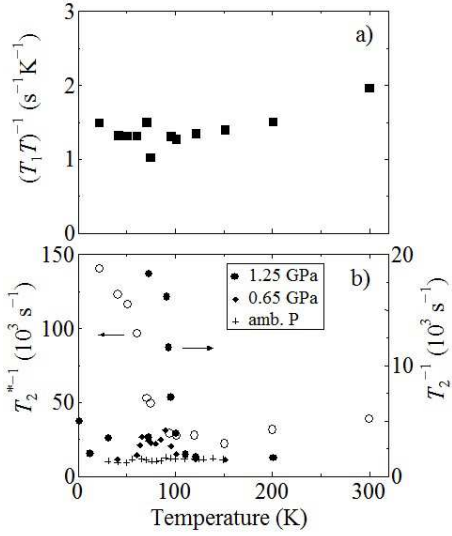} 		
\caption{
Temperature dependence of $^{77}$Se NMR relaxation rates at 1.25~GPa; 
a) $(T_1T)^{-1}$ and b) the transverse relaxation rate, $T_2^{-1}$ (closed circles), determined from echo decay curves.	
$T_2^{-1}$ at ambient pressure (crosses) and 0.65~GPa %
(closed diamonds)		%
shown in Fig.~\ref{fig:sum8kbar} e), are reproduced for comparison.		%
The inhomogeneous linewidth, $T_2^{*-1}$, 
determined as the square root of the second moment of the spectrum	%
at 1.25~GPa,  			%
is also plotted by open circles in b). 
}			
\label{fig:relax14kbar}
\end{center}
\end{figure}

We measured $^{77}$Se-NMR relaxation rates, $T_1^{-1}$ and $T_2^{-1}$ to investigate the spin/charge dynamics in the metallic state. 
The external filed, $H_0$,	was applied parallel to the $c^*$-axis, which is reported to give the shortest $T_1$\cite{zhang2005}.
Figure \ref{fig:relax14kbar}a) shows the temperature dependence of $(T_1T)^{-1}$.
A Korringa like relaxation behavior was observed in the whole measured temperature range. 
The value of $(T_1T)^{-1} \sim$ %
1.3~s$^{-1}$K$^{-1}$ 		%
is smaller than the values in the metallic states at lower pressures,
but this is %
again 				%
due to the reduction of the density of states at the Fermi level under pressure, as discussed before; 			
we have found that the Korringa constant, $(T_1TK^2)^{-1}$, is almost the same at ambient and under pressures.
We note that no anomalies were observed in the spectrum and the relaxations around 150~K, corresponding to transition V, 
as far as the $^{77}$Se-NMR properties are concerned. 

Contrary to the standard Korringa behaviour of the spin-lattice relaxation, $(T_1T)^{-1}$, 
a large enhancement of the transverse relaxation rate, $T_2^{-1}$,	
was observed in a narrow 	temperature range between 90~K and 70~K, as shown in Fig. \ref{fig:relax14kbar} b).
In fact, we found a remarkable reduction of echo signal intensities in this temperature %
region  		%
when we fixed the time interval, $\tau$', between $\pi /2$ and $\pi$ pulses to 50 $\mu$s.			%
The enhancement of $T_2^{-1}$, that is,  the broadening of the homogeneous width,	 
should be of the similar origin as those (double peaks) observed at lower pressure, 0.65~GPa, 
but now is found very much pronounced. 	
This may be partly because of the fact that the amplitude of the inhomogeneous fields is about twice as large at this pressure 
($\Delta K/K\sim0.3$ at 1.25~GPa, and 0.15 at 0.65~GPa). 	
For more qualitative comparison, the inhomogeneous width, $T_2^{*-1}$, 
determined as the square root of the second moment of the spectrum, is plotted by open circles in \ref{fig:relax14kbar} b). %
At high temperatures above $\sim$ 100~K, the main contribution of $T_2^{*-1}$ comes from the site dependence of the Knight shift. 
Below $\sim$ 70~K, 	the excess broadening becomes pronounced, as mentioned above. 		  %
If the present picture of the enhancement of $T_2^{-1}$ applies to the case, 		  %
the 		  %
maximum of the homogeneous width, $T_2^{-1}$ should be given, when $\langle\Delta\omega^2\rangle^{1/2}\tau_\mathrm{c}\sim$ 1, 
as $T_2^{-1})_\mathrm{max}\sim\langle\Delta\omega^2\rangle^{1/2}\tau_\mathrm{c}/2 \sim \langle\Delta\omega^2\rangle^{1/2}/2$,
where $\langle\Delta\omega^2\rangle^{1/2}$ should now be considered as the amplitude of the fluctuating component of the inhomogeneous width.\cite{chiba2004}  		
The observed maximum of $T_2^{-1}$ is actually about one half of $T_2^{*-1}$ around 70$\sim$90~K. 

Below $\sim$ 60~K, $T_2^{*-1}$ increases further up to 1.4 $\times 10^{5}$ s$^{-1}$$\sim$ 23~kHz at 1.7~K. 
This remarkable enhancement of $T_2^{*-1}$ just corresponds to those observed in Fig.~\ref{fig:vsp14kbar}~b) and c).  
We expect that this anomalous behaviour is related to transition III, observed in the higher pressure region of  the $T$-$P$ phase diagram. 
The $^{77}$Se NMR relaxations results suggest that,
below transition III,   
the dynamics of anions are almost frozen 
without any long range orderings of the dipolar orientations 
and the amplitude of the inhomogeneity $\Delta K/K$ increases remarkably probably because of some structural reasons.

%
\subsection{\label{sec:F-ap}$^{19}$F-NMR and anion dynamics at ambient pressure}	
\begin{figure}[htbp]
\begin{center}
\includegraphics[width=8cm
]{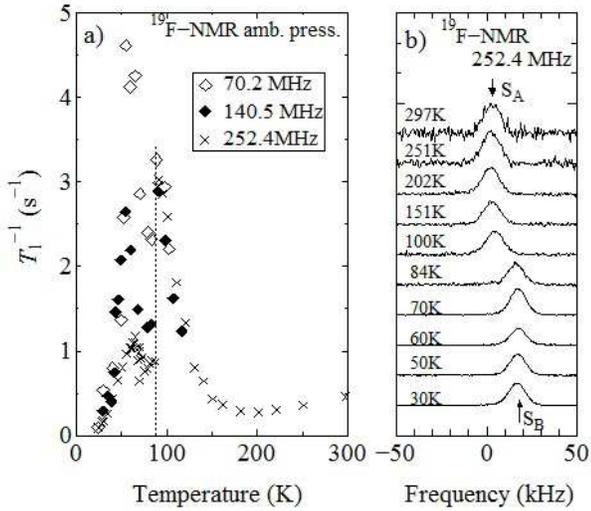} 
\caption{Temperature dependence of a)  $^{19}$F-NMR  $T_1^{-1}$  and b) spectrum at ambient pressure. 	%
The crosses, closed diamonds and open diamonds in a) are the data taken 							%
at 252~MHz, 140.5MHz and 70.2~MHz, respectively.   
The straight line indicates the temperature of transition I, 89~K. 		%
}
\label{fig:19F-ambP}
\end{center}
\end{figure} 

To clarify the role of the FSO$_3$ anion, we measured $^{19}$F-NMR spectrum and relaxation rate at ambient pressure, 
as well as under pressure of 0.45~GPa and 0.9~GPa.
Since there is no conduction electron at the anion site, the anion dynamics, that is, the rotational molecular motions, 
which may be of 3D manner at high temperatures or of a hindered type at low temperatures,  
are considered to affect the $^{19}$F-NMR properties.

The results of $^{19}$F-NMR relaxation and line shape measurements at ambient pressure 
are summarized in Fig. \ref{fig:19F-ambP}. 		
The spectral shape at 252~MHz  is a single line with a narrow width 
and does not show any change at temperatures above 100~K and below 84~K, 
but we found a small but appreciable shift of 15~kHz $\sim$ 70~ppm between the signals in the high and low temperature regions. 
We call tentatively the signals at high and low temperatures as signals S$_\mathrm{A}$ 
and S$_\mathrm{B}$, respectively. 	
We will discuss the assignment of these two signals later. 

The relaxation data of the ambient pressure, measured at different frequencies, are shown in a), 	%
by crosses (252~MHz), closed diamonds (140.5 MHz) and open diamonds (70.2~MHz). 
At each frequency, $T_1^{-1}$ increases with cooling down to 90 K, corresponding to transition I, 
at which a step-like drop of $T_1^{-1}$  is observed. 	
It does not show any appreciable frequency dependence above 90~K.
	
The high temperature data can be explained by the well-known BPP formula, which applies well for 
the 			%
relaxation 
due to random molecular motions, given as;\cite{bpp}
\begin{equation}
T_1^{-1}=(^{19}\gamma \Delta h)^2 \frac{\tau_\mathrm{c}^\mathrm{m}}{1+(\omega \tau_\mathrm{c}^\mathrm{m})^2} 
\end{equation}
where $\Delta h$, $\omega$ and $\tau_\mathrm{c}^\mathrm{m}$ are the amplitude of the fluctuation of the local magnetic field	
seen by the F-nuclei, $^{19}$F-NMR frequency and the correlation time of the local field fluctuation, respectively. 
Usually, the temperature dependence of $\tau_\mathrm{c}^\mathrm{m}$ is assumed to be 
$\tau_\mathrm{c}^\mathrm{m}=\tau_0 \exp(\Delta^\mathrm{m}/k_\mathrm{B}T)$ 
using a potential barrier for the molecular motions, $\Delta^\mathrm{m}$. 
This formula enables us to determine several important parameters characterizing the dynamics of the local fields 
from the observed relaxation behaviors.  		
The maximum of $T_1^{-1}$ is given at a temperature where $\omega \tau_\mathrm{c}^\mathrm{m}=1$, as  
$(T_1^{-1})_\mathrm{max} = (^{19}\gamma \Delta h)^2 /\omega$. 
The data at $\omega/2\pi$ = 252~MHz exhibit 
the 		%
maximum $\sim$ 3.0~s$^{-1}$ around $T$ $\sim$ 90~K, 
we obtained $\Delta h=$2.2 Oe.  
This is quite reasonable because the main contribution of the local fields at the 	fluorine site 
is considered to be the nuclear dipolar fields from the protons on the methyl groups at the edge of TMTSF molecule.
$\Delta^\mathrm{m}/k_\mathrm{B}$ was estimated as $\sim$ 670~K
using the formula for the high temperature limit  
($1 \gg (\omega \tau_\mathrm{c})^2$), $T_1^{-1}=(\gamma \Delta h)^2 \tau_\mathrm{0}\exp(\Delta^\mathrm{m}/k_\mathrm{B}T)$, 
which is frequency independent.

A step like drop of $T_1^{-1}$ at 90~K can be understood naturally as the breakdown of this BPP relaxation, 	%
since the 3D rotations at high temperatures should be forbidden by the anion order 
and the dynamics of anions should qualitatively change. 							%

Below 90~K, another peak of $T_1^{-1}$ appears around 70~K, which shows a strong frequency dependence; 
as the measured frequency lowered, the relaxation peak increases  in inverse proportion to frequency.
However, the frequency dependence cannot be explained by the BPP formula given above. 
Since almost all degrees of freedom are frozen in the AO state, 
only dynamics left giving rise to the fluctuations of the local fields at the F sites seem to be the hindered rotations of the methyl groups 
about their three-fold symmetry axes on the TMTSF molecule. 
We do not consider this further, since the effect of the methyl-group rotations on the electronic properties should be small.

\subsection{\label{sec:F-0.45}$^{19}$F-NMR and anion dynamics under pressure: 0.45~GPa}	
\begin{figure}[htbp]
\begin{center}
\includegraphics[width=8cm
]{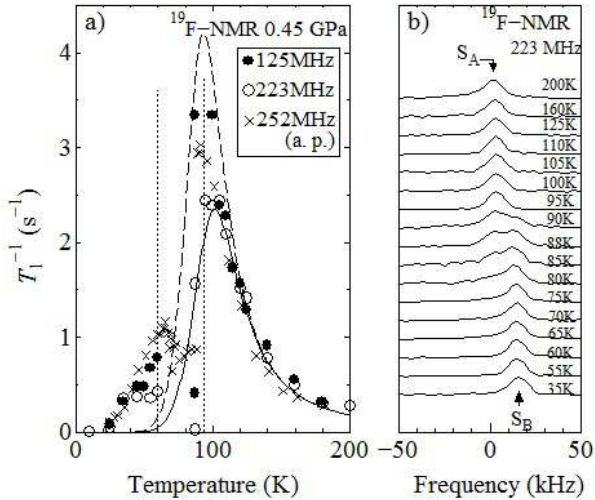}	 
\caption{Temperature dependence of a)  $^{19}$F-NMR  $T_1^{-1}$  and b) spectrum under pressure of 0.45~GPa.
The open and closed circles in a) are the data taken at 223~MHz and 125~MHz, respectively. 
The straight lines indicate the temperatures of transition I and II at 0.45~GPa. 		%
The curves show the fitting to the BPP formula given by eq. (1). 				%
Data at ambient pressure are also plotted for comparison by crosses.  			%
}
\label{fig:Vsp_4kbar}
\end{center}
\end{figure} 
The results 
of $^{19}$F-NMR relaxation and line shape measurements 
under pressure of 0.45~GPa		%
are summarized in the Fig. \ref{fig:Vsp_4kbar}. 
The relaxation behaviours are quite similar to those observed at ambient pressure. 
The observed temperature dependence above 90~K is very well explained as the BPP behaviour with an activation energy of
$\Delta^\mathrm{m}/k_\mathrm{B} \sim$ 670~K, which is almost the same as the value at ambient pressure.
A step-like drop of $T_1^{-1}$ is also observed at 90~K, corresponding to transition I,  
below which the 3D rotations of anion become strongly restricted. 
Considering that the (1/2 1/2 1/2) anion ordering occurs at 89 K at ambient pressure,\cite{moret1983}
similar anomalies can be expected at transition I under pressure. 
It was found that the additional 	relaxation seen below 90~K at ambient pressure was much suppressed under pressure. 
The methyl rotations or some other dynamics 
responsible for this additional relaxation 		%
should be markedly restricted by applying pressure. 

The spectral shape at 223~MHz  is a single line with a narrow width 
and does not show any change at temperatures above 90 K and below 60 K.  
The signals at the high and low temperatures correspond to the signals S$_\mathrm{A}$ 	%
and S$_\mathrm{B}$ observed at ambient pressure, respectively,	%
and they show a clear shift between them. 		%
In addition, both signals coexist in the intermediate temperature range where their relative intensity gradually changes.
The two characteristic temperatures correspond to transitions I and II, respectively.
The spectral shift should be attributed to the difference in the fluorine environments at high and low temperatures. 
It seems natural to attribute the signal S$_\mathrm{A}$ to the one from the anions rotating freely and the signal 
S$_\mathrm{B}$ to from the anions perfectly ordered. 
The coexistence of the two signals indicates that in the intermediated temperature region the anions 
are partially rotating and partially ordered. 
Actually, the relaxation rate, $T_1^{-1}$, for the signal S$_\mathrm{A}$ agrees well to the value expected for the BPP behavior, 
while the value for the signal S$_\mathrm{B}$ is much lower, as shown in Fig.~\ref{fig:Vsp_4kbar}. 
As temperature decreases, the fraction of rotating anions decreases and all anions are ordered below transition II.

Thus we believe that transition I corresponds to the temperature at which the freely rotating anion tetrahedrons 
form an orientational order but the F positions, or the orientations of permanent electric dipole,  are still random. 
Then the tetrahedrons make hindered rotations among four possible orientations for 	the F position, 
searching for a perfect ordering. 
The perfect anion order is 	realized at transition 	II. 
At ambient pressure, the ordering of tetrahedrons and that of electric dipoles take place  at the same transition at 89~K, 
as observed by the $X$-ray measurements. \cite{yamaura} 

It is noteworthy that the system remains metallic until the perfect anion order is realized.  
Furthermore, the unusual broadening of $^{77}$Se-NMR spectrum due to some inhomogeneous local fields 
appears only in the region where the anion orders remain imperfect. 
The $X$-ray study \cite{yamaura} observed that a superlattice appears at transition I 
but the oder parameter grows with cooling, until the system orders below transition II.
This is quite consistent with our observation.  

\subsection{\label{sec:F-0.9}$^{19}$F-NMR and anion dynamics under pressure: 0.9~GPa} 

\begin{figure}[htbp]
\begin{center}
\includegraphics[width=8cm
]{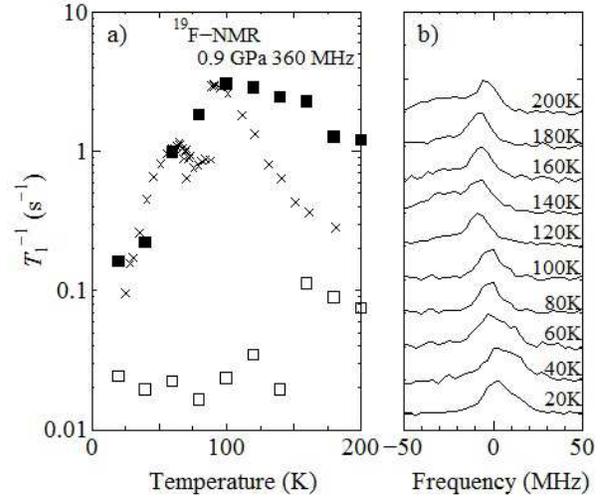} 
\caption{Temperature dependence of a) $^{19}$F-NMR  $T_1^{-1}$ and b) spectra at various temperatures under 0.9~GPa. 
Closed and open symbols in a) 	indicate the shorter and longer relaxation, respectively.  
Crosses indicate the data at ambient pressure. 	
}
\label{fig:Vsp_9kbar}
\end{center}
\end{figure} 

We applied a higher pressure of 0.9~GPa.
The results were shown in %
 Fig. \ref{fig:Vsp_9kbar}.
The $^{19}$F spectral shape and relaxation behaviors became much more complicated than those at 0.45~GPa.
The spectra were made of several components	and relaxation profiles became	non single exponential.
These suggest 	that the electronic structure has become more inhomogeneous	at this pressure.
There were observed no relaxation component which shows BPP-like behaviour any more. 
This means that the 3D rotations of anion may be %
entirely suppressed even at high temperatures, 
and instead the permanent dipoles may be randomly fixed with several different orientations. 
We suggest that transition V may be related to such a change in anion reorientations. 

It should be noted that the broadening of $^{77}$Se-NMR due to inhomogeneous local fields are more pronounced around this pressure, 
while the system remains metallic down to low temperatures.	

%
\section{\label{sec:conc}Concluding remarks}

\begin{figure}[htbp]
\begin{center}
\includegraphics[width=6cm
]{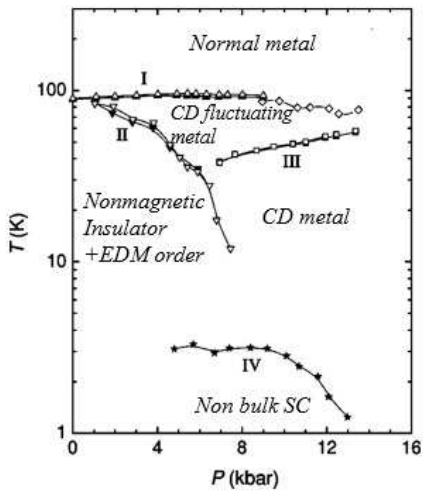}
\caption{
Temperature-Pressure phase diagram of (TMTSF)$_2$FSO$_3$, with electronic characteristics clarified by the present NMR measurements.  
Dotted lines are transitions V and VI, where  no NMR anomalies were observed. 
}
\label{fig:phaseDiagram}
\end{center}
\end{figure} 

We have performed $^{77}$Se and $^{19}$F NMR measurements to characterize the phases shown in the $T$-$P$ phase diagram of  (TMTSF)$_2$FSO$_3$ proposed on the basis of the transport anomalies. 
The $T$-$P$ phase diagram is reproduced in Fig.\ref{fig:phaseDiagram}, 		%
with the electronic characteristics clarified by the present NMR measurements. 

As we already reported,\cite{hiraki2003,satsukawa2005,takahashi2005}
the insulating state at ambient pressure below 90 K is nonmagnetic  with a finite spin gap, $\Delta/k_\mathrm{B}$ of $\sim$ 420 K. 
The $X$-ray analysis at ambient pressure claimed that at temperatures above 90~K, the FSO$_3^{-}$ anions are rotating in a 3D-manner.  	
The system behaves metallic in this temperature region. 
Below 90~K at ambient pressure, the tetrahedrons of anion, as well as the electric dipoles, 
are regularly ordered with the  wave number of (1/2 1/2 1/2).

By applying pressure, this transition is separated into two; transitions I and II.
We have proposed a picture as follows; 
the tetrahedrons of FSO$_{3}$ form an orientational order below transition I, 
while the electric dipoles are still randomly oriented and keep hopping among four possible directions 
with a characteristic time, $\tau_\mathrm{c}(T)$. 
When $\tau_\mathrm{c}(T)$ is long enough at low temperatures, the charge distribution within a TMTSF molecule is modified 
by the coupling with the electric dipoles of anion, which causes an inhomogeneous line broadening of $^{77}$Se-NMR spectrum, 
$\gamma\delta H \sim \gamma\Delta K H_{0}$, where $\Delta K = A\Delta\chi$ with the modulation of local susceptibility, $\Delta\chi$. 
At high temperatures, where  $\tau_\mathrm{c}(T)$ is so short to satisfy the condition, $\gamma\delta H\tau_\mathrm{c}(T) \ll 1$, 
the inhomogeneous local fields are averaged out and the line width is narrowed (motional narrowing). 
A large enhancement of homogeneous width is expected around the temperature where the condition,  
$\gamma\delta H\tau_\mathrm{c}(T) \sim 1$, is satisfied. 
This picture explains both 	the observed $^{77}$Se and $^{19}$F NMR properties consistently.	
The charge disproportionation which causes the NMR broadening in the present system is thus of  different nature 
from the one observed in  the other CD systems,  
(TMTTF)$_{2}$X (X= PF$_{6}$ and AsF$_{6}$) or $\theta$-(BEDT-TTF)$_{2}${\it M}Zn(SCN)$_{4}$. 
 
Below transition II, the electric dipoles form an orientational order, 
which should cause a periodic potential for the conduction electrons with the wave number of  (1/2 1/2 1/2), 
leading to a gap-opening on the Fermi level. 
The system becomes a nonmagnetic insulator with a finite spin-gap.	
The estimated value of the spin gap, $\Delta/k_\mathrm{B}$ is about 130 K 
under pressure of  0.65~GPa, which is much smaller than 420 K at ambient pressure\cite{satsukawa2005}.

It is noteworthy that the system remains metallic until the perfect ordering of electric dipoles are realized. 
This is strong contrast to the case of other TMTSF salts with symmetric tetrahedral anions, \textit{e. g.}, 
ReO$_{4}$, where the order of tetrahedrons could open a gap at the Fermi level leading to the metal-insulator transition. 
In the present FSO$_{3}$ salt, it seems that the randomness in the dipolar orientation strongly disturbs the formation of a periodic potential  
for conduction electrons.  

At 
 pressure of 1.25~GPa, no more perfect dipolar orderings, 		%
which could open gap on the FS, 		%
can be stabilized, 
so that transition II disappears and the system remains metallic down to the lowest temperature. 
Rapid dynamics of anions are expected in the phase region between transitions I and III, 
below which FSO$_{3}$ anions are frozen in a random orientation. 	
The intramolecular charge disproportionation is somewhat enhanced but remains in a moderate range less than 0.5 $\pm$ %
0.2. 		%
The present $^{77}$Se and $^{19}$F NMR results, however,  do not rule out the possibility of dipolar orderings with different wave number
without gap-opening at the FS, such as (0 0 1/2) or (0 1/2 0), suggested in the X-ray measurements.\cite{yamaura} 

In conclusion, the present measurements have confirmed the strong coupling of conduction electrons 
with the electric dipoles of the FSO$_3$ anions and succeeded to explain the natures of transitions I and II. 
It has been revealed that the electronic properties depend strongly on the dynamics of dipolar anions. 
We have suggested that transitions III and V should also be of some structural origins, 
but we need more detailed investigations on anion structures.

\begin{acknowledgment}

We thank Mr. R. Akagi for technical assistance. We also thank Prof. J. Yamaura (Tokyo Institute of Technology, Japan) 
for useful discussion about the structural point of view.
W. K. is supported by the NRF grants funded by the Korea Government (MSIP) (Grants No. 2015-001948). 
This work was supported by Grant-in-Aid for scientific research on priority area (No. 15073221) 
and on innovate area (No. 20110002) from ministry of education, culture, sports, science and technology, Japan.

\end{acknowledgment}


\begin{thebibliography}{99}

\bibitem{wudl1982}
F. Wudl, E. Aharon-Sharom, D. Nalewajek, J. V. Waszczak, W. M. Walsh, Jr., L. W. Rupp, Jr., P. M. Chaikin, R. Lacoe, M. Burns, T. O. Poehler, J. M. Williams, and M. A. Beno, J. Chem. Phys. \textbf{76}, 5497 (1982).

\bibitem{lacoe1983}
R. C. Lacoe, S. A. Wolf, P. M. Chaikin, F. Wudl, and E. Aharon-Shalom: Phys. Rev. \textbf{B27}, 1947 (1983)

\bibitem{gross1984}
F. Gross, H. Schwenk, K. Andres, F. Wudl, S. D. Cox and J. Brennan, 
Phys. Rev. \textbf{B30}, 1282 (1984)

\bibitem{jo2003}\label{bib:jo2003}
Y. J. Jo, E. S. Choi, Haeyong Kang, W. Kang, I. S. Seo, and O. H. Chung, Phys. Rev. \textbf{B67}, 014516 (2003) 


\bibitem{kang2003}
W. Kang, O. H. Chung, Y. J. Jo, Haeyong Kang, and I. S. Seo, 
Phys. Rev. \textbf{B68}, 073101 (2003)

\bibitem{hiraki2003}\label{bib:hiraki2003}
K. Hiraki, T. Nemoto, T. Takahashi, H. Kang, Y. J. Jo, W. Kang and O. H. Chung, Synth. Met. \textbf{135-136}, 691 (2003)

\bibitem{satsukawa2005}
H. Satsukawa, H. Kang, K. Hiraki, T. Takahashi, Y. J. Jo, W. Kang and O. H. Chung, Synth. Met. \textbf{153}, 417 (2005)


\bibitem{takahashi2005}
T. Takahashi, K. Hiraki, S. Moroto, N. Tajima, Y. Takano, Y. Kubo, H. Satsukawa, R. Chiba, H. M. Yamamoto, R. Kato and T. Naito, 
J. Phys. IV (France) \textbf{131}, 3 (2005)

\bibitem{nmrref}
External magnetic field was determined by $^{13}$C resonance frequency ($^{13}\nu_0$) of tetramethylsilan. 
$H_0$ was determined as $\frac{^{13}\nu_{0}}{^{13}\gamma}$, where $^{13}\gamma$ is the gyromagnetic ratio of $^{13}$C, 10.7054 MHz/T.

\bibitem{dimension}
The diameter of the sample space of the cryostat is 75mm (TeslatronH type superconducting magnet with VTI: Oxford instruments inc.).

\bibitem{Daphne7373}
K. Murata, H. Yoshino, H. O. Yadav, Y. Honda and N. Shirakawa, Rev. Sci. Instrum. \textbf{68}, 2490 (1997)

\bibitem{zhang2005}
F. Zhang, Y. Kurosaki, J. Shinagawa, B. Alavi, and S. E. Brown, Phys. Rev. \textbf{B72}, 060501 (2005)

\bibitem{yamaura}
J. Yamaura, unpublished

\bibitem{AO_gap}
H. Schwnek, K. Andres and F. Wudl, Phys. Rev. \textbf{B29}, 500 (1984)

\bibitem{chiba2004}
R. Chiba, K. Hiraki, T. Takahashi, H. M. Yamamoto, and T. Nakamura, 
Phys. Rev. Lett. \textbf{93}, 216405 (2004)


\bibitem{hiraki2007}
K. Hiraki, H. Mayaffre, M. Horvatic, C. Berthier, T. Yamaguchi, S. Uji, H. Tanaka, A. Kobayashi, H. Kobayashi and T. Takahashi, 
J. Phys. Soc. Jpn. \textbf{76}, 124708 (2007)

\bibitem{hiraki2010}
K. Hiraki, M. Kitahara, T. Takahashi, H. Mayaffre, M. Horvatic, C. Berthier, S. Uji, H. Tanaka, B. Zhou, A. Kobayashi and H. Kobayashi, 
J. Phys. Soc. Jpn. \textbf{79}, 074711 (2010)

\bibitem{brown2002}
F. Zamborszky, W. Yu, W. Raas, S. E. Brown, B. Alavi, C. A. Merlic and A. Baur, 
Phys. Rev. \textbf{B66}, 081103(R) (2002)

\bibitem{fujiyama2006}
S. Fujiyama and T. Nakamura, J. Phys. Soc. Jpn. \textbf{75}, 014705 (2006)

\bibitem{hirose2010}
S. Hirose, A. Kawamoto, N. Matsunaga, K. Nomura, K. Yamamoto, and K. Yakushi, 
Phys. Rev. \textbf{B81} 205107 (2010)

\bibitem{bpp}N. Bloembergen, E. M. Purcell, and R. V. Pound, Phys. Rev. \textbf{73}, 679 (1948).

\bibitem{moret1983}
R. Moret, J. P. Pouget, R. Com\`{e}s and K. Bechgaard, J. Physique colloque \textbf{44}, C3-957 (1984)

\end{thebibliography}
\end{document}